\documentclass[reqno,a4,11pt]{amsart}
\usepackage{amsmath,amssymb,tabularx,setspace,color,multirow,graphics,graphicx}
\newtheorem{theorem}{Theorem}

\newtheorem{definition}{Definition}
\newtheorem{remark}{Remark}
\newtheorem{corollary}{Corollary}
\makeatletter
\renewcommand\section{\@startsection {section}{1}{\z@}%
                                   {-3.5ex \@plus -1ex \@minus -.2ex}%
                                   {2.3ex \@plus.2ex}%
                                   {\normalfont\large\bfseries}}
\makeatother

\newtheorem{Corollary}{Corollary}

\newtheorem{Theorem}{Theorem}

\newtheorem{Example}{Example}

\begin{document}
\doublespace
\vspace{-0.3in}
\title[]{ Jackknife empirical likelihood ratio test for testing the equality of  semivariance }
\author[]%
{   S\lowercase{aparya}  S\lowercase{uresh\textsuperscript{a} and} S\lowercase{udheesh} K. K\lowercase{attumannil,\textsuperscript{b}  }
\\
 \lowercase{\textsuperscript{a}}I\lowercase{ndian} I\lowercase{nstitute of } M\lowercase{anagement}, K\lowercase{ozhikode}, I\lowercase{ndia},\\ \lowercase{\textsuperscript{b}}I\lowercase{ndian} S\lowercase{tatistical} I\lowercase{nstitute},
  C\lowercase{hennai}, I\lowercase{ndia}.}
\thanks{{$^{\dag}$}{Corresponding E-mail: \tt skkattu@isichennai.res.in}}
\maketitle
\vspace{-0.2in}

\begin{abstract}
 Semivariance is a measure of the dispersion of all observations that fall above the mean or target value of a random variable and it plays an important role in life-length,  actuarial and income studies. In this paper, we develop a new non-parametric test for equality of upper semi-variance. We use the U-statistic theory to derive the test statistic and then study the asymptotic properties of the test statistic. We also develop a jackknife empirical likelihood (JEL) ratio test for equality of upper Semivariance.  Extensive Monte Carlo simulation studies are carried out to validate the performance of the proposed JEL-based test. We illustrate the test procedure using real data.  
\end{abstract}
\vspace{-0.1in}
\section{Introduction}
Semivariance is a measure of the dispersion of all observations that fall above the mean or target value of a random variable. Semivariance is an average of the squared deviations of values that are more than the target value or the mean. Semivariance is similar to variance; however, it only considers the observations above the target value or the mean. For a non-negative random variable $X$, we give the definition of semivariance which is used in the sequel.
\begin{definition}\label{pmr}
	 Consider a non-negative random variable $X$  having cumulative distribution function  $F(.)$.  The target semivariance of  $X$ above a target value $t$ is defined as follows:
  \begin{equation}\label{sv}
      \beta_X(t)=\int_{t}^{\infty}(x-t)^2 dF(x).
  \end{equation}
\end{definition}
\noindent A generalised version of $ \beta_X(t)$ with a given power $r>0$ in the above expression are  called stop-loss moments.

The stop-loss moments play an important role in life-length,  actuarial and income studies. One example is the random variable $(X-t)_+$, which is employed to define partial moments that have significance in the analysis of personal earnings. Let $X$ denote an individual's income and $t$ represent the taxable income level. In this context, $(X-t)_+$ represents the taxable income. Income below the tax exemption threshold $t$ does not impact tax calculations and is thus considered as zero. Hence, the examination of these higher-order moments is valuable in analyzing observations that surpass a certain threshold without truncating the distribution at $t$. The term $(X-t)_+$ is frequently employed in actuarial research to represent the financial losses incurred by an insurance firm. The expression  $(X-t)_+$ can be used as a metric for evaluating the risks undertaken by the company and is widely recognized as a standard indicator of the risk level connected with $X$ (Cheng and Pai, 2003).

 Nair et al. (2013) studied the properties of partial moments using quantile functions. As an application to reliability modeling, they characterized certain lifetime quantile function models. The stop loss function is employed in the modeling of the duration of inactivity resulting from the failure of a component, provided that its lifespan is less than or equal to $t$.      Kundu and Sarkar (2017) defined certain continuous distributions by analyzing the conditional partial moments of inactive time.

 The importance of partial moments has been explored in bivariate setup as well. Sunoj and Vipin (2019) studied conditional partial moments and obtained some characterization results for bivariate life distribution.  Nair et al. (2020) conducted a comprehensive investigation on the bivariate stop-loss transform within the framework of reliability analysis. Nair et al. (2020) also studied some properties of higher-degree bivariate partial moments by extending the ideas of characterizing a univariate identity of partial moments. Vipin and Sunoj (2021) studied the partial moments in bivariate setup using copula functions. Recently, Vipin et al. (2023) studied the properties of partial moments under biased sampling. Further, Nawrocki and Viole (2023) showed that semivariance exhibits strong boundary equivalence to Chebychev's inequality and also captures significant skewness and leptokurtic behaviour making it an ideal tool for studying asset distributions.

Motivated by the use of semivariance, in this work, we propose a test for assessing the equality of semivariance between two different random variables. This test enables the comparison of variance in taxable income of two different entities in income studies or the analysis of the risk undertaken by two different insurance companies in actuarial studies. Section 2 details the derivation of the test statistics using the theory of U-statistics and its asymptotic properties. Due to difficulties in the implementation of the normal-based test statistic, we also proposed a jackknife empirical likelihood ratio test, which is distribution-free.  In section 3, a simulation study of the proposed test is carried out for different scenarios. Section 4, looks at the performance of the proposed test on real data sets and finally, the conclusion along with some related open problems is given in Section 5.

\section{Testing equality of the semivariance}
We develop a test for equality of the semivariance. Consider two non-negative random variables $X$ and $Y$    having cumulative distribution functions $F(.)$ and $G(.)$, respectively.  Denote $\beta_Y(t)$ as the semivariance of  $Y$ as defined in equation (\ref{sv}).  Let  $X_{1}, ...,X_{n_1}$ and $Y_{1}, ...,Y_{n_2}$ be  random samples of size $n_1$ and $n_2$ from $F$ and $G$, respectively. We are interested in  testing the null hypothesis
$$H_{0}: \beta_X(t)=\beta_Y(t),\,\forall  t>0$$
against the alternatives hypothesis $$ H_1: \beta_X(t)\ne \beta_Y(t)\,\text{ for some } t>0.$$
 For testing the above hypothesis, first we define a departure measure that discriminates between the null and the alternative hypotheses. Hence, we consider a measure of departure $ \Delta(F,G)$ given by
\begin{eqnarray}\label{deltam}
\Delta(F,G)&=&\int_{0}^{\infty}(\beta_X(t)-\beta_Y(t))(dF(t)+dG(t))\nonumber\\&=&
\int_{0}^{\infty}\beta_X(t)dF(t)+\int_{0}^{\infty}\beta_X(t)dG(t)-\int_{0}^{\infty}\beta_Y(t)dF(t)\nonumber\\&&-\int_{0}^{\infty}\beta_Y(t)dG(t)\nonumber\\&=&\Delta_1+\Delta_2-\Delta_3-\Delta_4.
\end{eqnarray}Clearly, $\Delta(F,G)$ is zero under  $H_0$ and non-zero  under $H_1$. Hence $\Delta(F,G)$  can be considered as a measure of departure from the null hypothesis $H_0$ towards the alternative hypothesis $H_1$. Based on the departure measure $\Delta(F,G)$ we first we develop a normal based test.
\subsection{U-statistics based test}
 We use the theory of U-statistics to derive an estimator of the proposed departure measure. Thus,  we simplify $\Delta(F,G)$ in terms of the expectation of the function of random variables.

Consider,
\begin{eqnarray}\label{delta1}
\Delta_1&=&\int_{0}^{\infty}\int_{t}^{\infty}(x-t)^2dF(x)dF(t)\nonumber\\
&=&\int_{0}^{\infty}\int_{t}^{\infty}x^2dF(x)dF(t)+\int_{0}^{\infty}\int_{t}^{\infty}t^2dF(x)dF(t)\nonumber\\&&-2\int_{0}^{\infty}\int_{t}^{\infty}xtdF(x)dF(t)\nonumber\\
&=&\int_{0}^{\infty}x^2\int_{0}^{x}dF(t)dF(x)+\int_{0}^{\infty}t^2\int_{t}^{\infty}dF(x)dF(t)\nonumber\\&&-2\int_{0}^{\infty}\int_{t}^{\infty}xtdF(x)dF(t)
\nonumber\\
&=&\int_{0}^{\infty}x^2F(x)dF(x)+\int_{0}^{\infty}t^2 \bar F(t)dF(t)-2\int_{0}^{\infty}\int_{t}^{\infty}xtdF(x)dF(t)
\nonumber\\
&=&\int_{0}^{\infty}x^2dF(x)-2\int_{0}^{\infty}\int_{t}^{\infty}xtdF(x)dF(t)
\nonumber\\
&=&\int_{0}^{\infty}x^2dF(x)-2\int_{0}^{\infty}\int_{0}^{\infty}xtI(x>t)dF(x)dF(t)
\nonumber\\
&=&E(X_1^2)-2E\big(X_1X_2I(X_1>X_2)\big).\nonumber
\end{eqnarray}
On similar lines as above, we obtain
\begin{eqnarray}\label{delta3}
\Delta_4&=&\int_{0}^{\infty}\int_{t}^{\infty}(x-t)^2dG(x)dG(t)\nonumber\\
&=&E(Y_1^2)-2E\big(Y_1Y_2I(Y_1>Y_2)\big).
\end{eqnarray}
Now, consider
\begin{eqnarray}\label{delta2}
\Delta_2&=&\int_{0}^{\infty}\int_{t}^{\infty}(x-t)^2dF(x)dG(t)\nonumber\\
&=&\int_{0}^{\infty}\int_{t}^{\infty}x^2dF(x)dG(t)+\int_{0}^{\infty}\int_{t}^{\infty}t^2dF(x)dG(t)\nonumber
\\&&-2\int_{0}^{\infty}\int_{t}^{\infty}xtdF(x)dG(t).\nonumber
\end{eqnarray}Changing the order of integration, we obtain
\begin{eqnarray}\label{delta2}
\Delta_2&=&\int_{0}^{\infty}x^2\int_{0}^{x}dG(t)dF(x)+\int_{0}^{\infty}t^2\int_{t}^{\infty}dF(x)dG(t)\nonumber\\&&-2\int_{0}^{\infty}\int_{t}^{\infty}xtdF(x)dG(t)
\nonumber\\
&=&\int_{0}^{\infty}x^2G(x)dF(x)+\int_{0}^{\infty}t^2 \bar F(t)dG(t)-2\int_{0}^{\infty}\int_{t}^{\infty}xtdF(x)dG(t)
\nonumber\\
&=&\int_{0}^{\infty}x^2G(x)dF(x)+\int_{0}^{\infty}t^2 \bar F(t)dG(t)-2E\big(X_1Y_1I(X_1>Y_1)\big)
\nonumber\\
&=&E\big(X_1^2I(X_1>Y_1)\big)+E\big(Y_1^2\big)-E\big(Y_1^2I(Y_1>X_1)\big)\nonumber\\&&-2E\big(X_1Y_1I(X_1>Y_1)\big).
\end{eqnarray}
Similar lines as above, we obtain
\begin{eqnarray}\label{delta4}
\Delta_3&=&\int_{0}^{\infty}\int_{t}^{\infty}(x-t)^2dG(x)dF(t)\nonumber\\
&=&\int_{0}^{\infty}x^2F(x)dG(x)+\int_{0}^{\infty}t^2 \bar G(t)dF(t)-2E\big(X_1Y_1I(Y_1>X_1)\big)
\nonumber\\
&=&E\big(Y_1^2I(Y_1>X_1)\big)+E\big(X_1^2\big)-E\big(X_1^2I(X_1>Y_1)\big)\nonumber\\&&-2E\big(X_1Y_1I(X_1>Y_1)\big).
\end{eqnarray}
Substituting equations (\ref{delta1}) - (\ref{delta4}) in (\ref{deltam}), we obtain
\begin{eqnarray}\label{deltafinal}
\Delta(F,G)&=&2E\big(Y_1Y_2I(Y_1>Y_2)\big)-2E\big(X_1X_2I(X_1>X_2)\big)
\nonumber\\&&+2E\big(X_1Y_1I(Y_1>X_1)\big)-2E\big(X_1Y_1I(X_1>Y_1)\big)\nonumber\\&&+2E\big(X_1^2I(X_1>Y_1)\big)
-2E\big(Y_1^2I(Y_1>X_1)\big).
\end{eqnarray}

In view of the representation (\ref{deltafinal}),  we propose the test  based on  U-statistics and it is given by
\begin{equation}
 \widehat{\Delta} =\left(\frac{2}{n_1(n_1-1)}\right)\left(\frac{2}{n_2(n_2-1)}\right)  \sum_{i=1}^{n_1-1}\sum_{j=i+1}^{n_1}\sum_{k=1}^{n_2-1}\sum_{l=k+1}^{n_2}
 h(X_i,X_j,Y_k,Y_l),
\end{equation}where
\begin{small}
\begin{eqnarray*}
h(X_1,X_2,Y_1,Y_2)&=&Y_1Y_2I(Y_1>Y_2)+Y_1Y_2I(Y_2>Y_1)-X_1X_2I(X_1>X_2)\\&&-X_1X_2I(X_2>X_1)
+X_1Y_1I(Y_1>X_1)+X_2Y_2I(Y_2>X_2)\\&&-X_1Y_1I(X_1>Y_1)-X_2Y_2I(X_2>Y_2)+
X_1^2I(X_1>Y_1)\\&&+X_2^2I(X_2>Y_2)
-Y_1^2I(Y_1>X_1)-Y_2^2I(Y_2>X_2).
\end{eqnarray*}
\end{small}
By definition of U-statistics, $ \widehat{\Delta}$ is unbiased estimator of  $\Delta(F,G)$.
 We reject the null hypothesis $H_0$ against the alternative hypothesis  $H_1$ for a large value of $\widehat{\Delta}$. We find a critical region of the test based on the asymptotic distribution of  $\widehat{\Delta}.$

 Next, we find the asymptotic distribution of $\widehat{\Delta} $. Using the central limit theorem for U-statistics we have the asymptotic normality of $ \widehat{\Delta}$  and we state it as the following theorem (Lee, 2019).\vspace{-0.1in}
\begin{theorem}  Let $n=n_1+n_2$ and $p=\lim_{n \rightarrow \infty} \frac{n_1}{n}$,  $0< p <1.$
  As $n\rightarrow \infty$,  $\sqrt{n}(\widehat{\Delta}-\Delta (F,G))$ converges in distribution to a normal random variable with mean zero and variance $\sigma^2$, where $\sigma^2$ is given by
   \begin{equation}\label{var}
\sigma^{2}= \frac{4}{p}\sigma_{10}^2+\frac{4}{1-p}\sigma_{01}^2,
\end{equation}with
$$\sigma^2_{10} =Cov (h(X_1,X_2,Y_1,Y_2),h(X_1,X_3,Y_1,Y_2))$$
\mbox{and}
$$\sigma^2_{01} =Cov (h(X_1,X_2,Y_1,Y_2),h(X_1,X_2,Y_1,Y_3)).$$
\end{theorem}\vspace{-0.1in}
\noindent Under $H_0$, we have $\Delta(F,G) =0$.  Hence we have the following corollary.\vspace{-0.1in}
\begin{corollary}
 Under $H_0$, as $n\rightarrow \infty$,  $\sqrt{n}\widehat{\Delta}$ converges in distribution to a normal random variable with mean zero and variance $\sigma_0^2$, where $\sigma_0^2$ is given by
\begin{equation}\label{varnull}
\sigma_{0}^{2}= \frac{4}{p(1-p)}\sigma_{10}^2
\end{equation}
with
\begin{equation}\label{nullvarf}
 \sigma_{10}^2=Var\left(X^2F(X)-2X\int_{0}^{X} ydF(y)-\int_{X}^{\infty}y^2dF(y)\right).
\end{equation}
\end{corollary}
\noindent{\bf Proof:} Under $H_0$, the distribution of $X$ and $Y$ are same. Hence the variance expression given in eqaution  (\ref{var}) reduces to (\ref{varnull}).   Using Theorem 1 of Chapter 2 of Lee (2019), we have
\begin{eqnarray}\label{eq12}
\sigma^2_{10}& =&Cov (h(X_1,X_2,Y_1,Y_2),h(X_1,X_3,Y_1,Y_2)) \nonumber\\
&=&Var(E(h(X_1,X_2,Y_1,Y_2)|X_1)).
\end{eqnarray}
Consider
\begin{small}
\begin{eqnarray*}
E(h(X_1,X_2,Y_1,Y_2)|X_1=x)&=&E\Bigg(\begin{small}Y_1Y_2I(Y_1>Y_2)+Y_1Y_2I(Y_2>Y_1)-xX_2I(x>X_2)\end{small}\\&&-xX_2I(X_2>x)
+xY_1I(Y_1>x)+X_2Y_2I(Y_2>X_2)\\&&-xY_1I(x>Y_1)-X_2Y_2I(X_2>Y_2)+
x^2I(x>Y_1)\\&&+X_2^2I(X_2>Y_2)
-Y_1^2I(Y_1>x)-Y_2^2I(Y_2>X_2)\Bigg).
\end{eqnarray*}
\end{small}
\noindent Denote $k$ as a real constant. Then, the above equation can be written as
\begin{small}
\begin{eqnarray*}
E(h(X_1,X_2,Y_1,Y_2)|X_1=x)&=&-x\int_{0}^{\infty} yI(x>y)dF(y)-x\int_{0}^{\infty} yI(y>x)dF(y)
\\&&+x\int_{0}^{\infty}yI(y>x)dG(y) -x\int_{0}^{\infty}yI(x>y)dG(y)\\&&+
x^2P(x>Y_1)-\int_{0}^{\infty}y^2I(y>x)dG(y)+k,
\end{eqnarray*}
\end{small}which reduces to
\begin{eqnarray*}
E(h(X_1,X_2,Y_1,Y_2)|X_1=x)=x^2G(x)-2x\int_{0}^{x} ydG(y)-\int_{x}^{\infty}y^2dG(y)+k.
\end{eqnarray*}Or
\begin{eqnarray*}
E(h(X_1,X_2,Y_1,Y_2)|X_1=x)=x^2F(x)-2x\int_{0}^{x} ydF(y)-\int_{x}^{\infty}y^2dF(y)+k.
\end{eqnarray*}Hence, from (\ref{eq12}),  we obtain the variance expression specified in equation (\ref{nullvarf}).

An asymptotic critical region of the normal-based test can be obtained using Corollary 1. Let $S^2$ be a consistent estimator of the asymptotic null variance $\sigma_{0}^2$. We reject the  null hypothesis $H_{0}$ against the alternative hypothesis $H_{1}$ at an approximate  significance level $\alpha$, if
\begin{equation*}
 \frac{ \sqrt{n} |\widehat{\Delta}| }{S}>Z_{\alpha/2},
  \end{equation*}
where $Z_{\alpha}$ is the upper $\alpha$-percentile point of the standard normal distribution.

Obtaining a reliable estimate of the asymptotic null variance $\sigma_{0}^2$ is challenging, making it infeasible to implement an asymptotic test based on the normal distribution. This motivated us to develop a jackknife empirical likelihood (JEL) ratio test for testing the equality of semivariance.

\vspace{-0.2in}
\subsection{JEL ratio test}
Thomas and Grunkemier (1975) introduced the idea of empirical likelihood to obtain the confidence interval for survival probabilities under right censoring. Owen (1988, 1990) extended this concept of empirical likelihood to a general methodology. Computationally, empirical likelihood involves maximizing non-parametric likelihood supported on the data subject to some constraints. If the constraints become non-linear, it poses a computational difficulty to evaluate the likelihood. Further, it becomes increasingly difficult as $n$ gets large. To overcome this problem, Jing et al.  (2009) introduced the jackknife empirical likelihood method to find a confidence interval for desired parametric function. They illustrated the proposed method using one sample and two sample U-statistics. This approach is widely accepted among researchers as it combines the effectiveness of the likelihood approach and the jackknife technique.

Next, we discuss the jackknife empirical likelihood method for testing  the equality of semivariance. Accordingly, first we need find the jackknife pseudo-values. For this purpose, we express the test statistics $\hat{\Delta}$ as

\begin{eqnarray} \nonumber
    \widehat{\Delta} &=&\frac{1}{\binom{n_1}{2} \binom{n_2}{2}} \sum_{i<j=1}^{n_1}\sum_{k<l=1}^{n_2} h_1 (X_i,X_j,Y_k,Y_l)\\ \nonumber
    &=& T_1 (X_1,X_2,\cdots,X_n,Y_1,Y_2,\cdots,Y_n).
\end{eqnarray}

 Let $n=n_1+n_2$ be the total sample size and define $Z_i$ as
 $$Z_i =\begin{cases}
     X_i \mbox{  if  } i= 1, \cdots, n_1\\
     Y_j  \mbox{  if  } j= n_1+1, \cdots, n.\\
 \end{cases}$$
 Then, we can rewrite $\widehat{\Delta}$ as
 $$ \widehat{\Delta} = T_1(Z_1, \cdots, Z_n).$$
 Let $\nu_i$ be the jackknife pseudo-values defined as follows:
$$\nu_i= nT_1 -(n-1) T_{1,i},  \hspace{ 0.25 cm} i =1,\cdots,n,$$
 where $T_{1,i}$ is the value of $\widehat{\Delta}$ obtained using $(n-1)$ observation $Z_1,\cdots,Z_{i-1},$ $Z_{i+1}, \cdots,Z_n$.
 It can be observed that
 $$\hat{\Delta} = \frac{1}{n}\sum_{i=1}^{n} \nu_i.$$
It can be shown that $\nu_i, i =1,\cdots, n$ are asymptotically independent random variables. Hence, the jackknife estimators $\widehat{\Delta}$ is the average of asymptotically independent random variables $\nu_i$.

 Let $p=(p_1,\ldots,p_n)$ be a probability vector assigning probability $p_i$ to each $v_i$. Note that the objective function $\prod_{i=1}^{n}p_i$ subject to the constraints $\sum_{i=1}^{n}p_i=1$ attain its maximum value $n^{-n}$ when all $p_i=1/n$.

Thus, the jackknife empirical likelihood ratio for testing for the equality of semivariance   based on the departure measure ${\Delta}$ is  given by

$$R(\Delta)=\max\Big\{\prod_{i=1}^{n} np_i:\,\sum_{i=1}^{n}p_i=1,\,\,\sum_{i=1}^{n}p_i\nu_i=0\Big\}.$$
Using the Lagrange multiplier method, we obtain
$$p_i=\frac{1}{n}\frac{1}{(1+\lambda \nu_i)},$$
where $\lambda$ satisfies $$\frac{1}{n}\sum_{i=1}^{n}\frac{\nu_i}{1+\lambda \nu_i}=0.$$
Thus the jackknife empirical log-likelihood ratio is given by $$\log R(\Delta)=-\sum\log(1+\lambda \nu_i).$$
In view of Theorem 1 above, using Theorem 2 of Jing et al. (2009), we have the following result as an analog of Wilk's theorem on the limiting distribution of the jackknife empirical log-likelihood ratio.\vspace{-0.1in}
\begin{theorem}  \label{thmjel}
Suppose  $E(h^2(X_1,X_2,Y_1,Y_2))<\infty$, $\sigma^2_{10} $ and $\sigma^2_{01}>0$. Assume that $0< \liminf \frac{n_1}{n_2}< \limsup \frac{n_1}{n_2} < \infty $. Under $H_0$, as $n\rightarrow\infty$,  $-2\log R(\Delta)$ converges in distribution to a $\chi^2$ random variable with one degree of freedom.
\end{theorem}\vspace{-0.1in}
Using  Theorem \ref{thmjel},  we obtain the critical region of the JEL ratio test.  We  reject the null hypothesis $H_0$ against the alternatives hypothesis $H_1$ at a  significance level $\alpha$, if
 \begin{equation}
 	-2 \log R(\Delta)> \chi^2_{1,\alpha},
 \end{equation}	
where $\chi^2_{1,\alpha}$ is the upper $\alpha$-percentile point of the $\chi^2$ distribution with one degree of freedom.
\vspace{-0.1in}
\section{Simulation Study}\vspace{-0.1in}
In this section, we carried out a Monte Carlo simulation study to assess the finite sample performance of the proposed JEL ratio  test. The simulation is done ten thousand times using different sample sizes 20, 40, 60, 80 and 100. Simulation is done using R software.

 We carried out the simulation study by taking the probability distributions of the $X$ and $Y$ as exponential, Pareto and lognormal distributions. For finding the empirical type 1 error, we assumed that both $X$ and $Y$ follow the same distribution. The results from the simulation study are given in Tables \ref{tab:ex}, \ref{tab:par}, and  \ref{tab:lnorm}.  From these tables, we observe that the empirical type 1 error converges to the significance level $\alpha$ as the sample size $n$ increases. Contrary to the usual behavior, for lognormal, we can observe that empirical type 1 error is underestimated for lower sample sizes. However, we can see that as the sample sizes increase, the empirical type 1 error converges to the significance level $\alpha$.

 To find the empirical power of the proposed JEL test, we assumed different distributions for $X$ and $Y$. We considered three different scenarios:
 \begin{itemize}
     \item  Scenario 1: $X$ following a lognormal distribution with parameters $(\mu,\sigma)$ and $Y$ following an exponential distribution with parameter  $\lambda$.
     \item Scenario 2: $X$ following a lognormal distribution with parameters $(\mu,\sigma)$ and $Y$ following a Pareto distribution  with parameter $\alpha$.
     \item Scenario 3: $X$ following  an exponential distribution with parameter $\lambda$ and $Y$ following a Pareto distribution  with parameter $\alpha$.
 \end{itemize}

 The simulation results of each scenario considered above are given in Tables \ref{tab:sc1}, \ref{tab:sc2} and \ref{tab:sc3} respectively. From these tables, we can see that the empirical power of the test converges to one as the sample size increases. In Scenarios 2 and 3, where one of the random variables is assumed to be Pareto, the empirical power attained one even at a small sample size $n=20$.
\begin{table}[]
\caption{Empirical Type 1 error for exponential distribution at 5\% significance level.  }
    \centering
    \begin{tabular} {cccccccccc}
 \hline
 & $n$ & $\lambda=2$ & $\lambda=3$& $\lambda=4$ \\
\hline

   &20 & 0.074& 0.083 & 0.075   \\
    &40 & 0.069& 0.071 & 0.073\\
    &60 & 0.063 & 0.057& 0.066 \\
    &80 & 0.054& 0.054& 0.053\\
    &100 & 0.051  & 0.052& 0.050\\
    \hline
    \end{tabular}

    \label{tab:ex}
\end{table}

\begin{table}[]
\caption{Empirical Type 1 error for Pareto distribution at 5\% significance level.  }
    \centering
    \begin{tabular} {cccccccccc}
 \hline
 & $n$ & $\alpha=2$ & $\alpha=3$& $\alpha=4$ \\
\hline

   &20 & 0.0.097& 0.093 & 0.096   \\
    &40 & 0.065& 0.076 & 0.068\\
    &60 & 0.061 & 0.061& 0.059 \\
    &80 & 0.054& 0.057& 0.052\\
    &100 & 0.051  & 0.050& 0.050\\
    \hline
    \end{tabular}

    \label{tab:par}
\end{table}

\begin{table}[]
\caption{Empirical Type 1 error for lognormal distribution at 5\% significance level.  }
    \centering
    \begin{tabular} {cccccccccc}
 \hline
 & $n$ & $(\mu,\sigma)= (0,1)$ & $(\mu,\sigma)= (0,2)$ & $(\mu,\sigma)= (1,2)$\\
\hline

   &20 & 0.012& 0.023 & 0.015   \\
    &40 & 0.031& 0.031 & 0.024\\
    &60 & 0.040 & 0.040& 0.035 \\
    &80 & 0.056& 0.057& 0.047\\
    &100 & 0.050  & 0.050& 0.052\\
    \hline
    \end{tabular}

    \label{tab:lnorm}
\end{table}

\begin{table}[]
\caption{Empirical power for Scenario 1 at 5 \% significance level}
    \centering
    \scalebox{0.9}{
    \begin{tabular} {cccccccccc}
    \hline
   $n$ &  ($\lambda =2$,  $(\mu,\sigma)= (0,1)$) &($\lambda =3$,  $(\mu,\sigma)= (0,1)$) & ($\lambda =2$,  $(\mu,\sigma)= (1,2)$)\\
\hline

   20 & 0.901& 0.913 & 0.907 \\
    40 & 0.978&0.967 & 0.973 \\
    60 & 0.990&0.994 &0.989 \\
    80 & 0.997& 1.000 &0.999\\
    100 & 1.000 &1.000& 1.000\\

    \hline
    \end{tabular}

    \label{tab:sc1}}
\end{table}

\begin{table}[]
\caption{Empirical power for Scenario 2  at 5 \% significance level}
    \centering
      \scalebox{0.9}{
    \begin{tabular} {cccccccccc}
    \hline
   $n$ & ($\alpha =2$,  $(\mu,\sigma)= (0,1)$) &($\alpha =3$,  $(\mu,\sigma)= (0,1)$) & ($\alpha =2$,  $(\mu,\sigma)= (1,2)$)\\ \\
\hline

   20 & 1.000& 1.000 &1.000\\
    40 & 1.000 & 1.000 &1.000\\
    60 & 1.000& 1.000 &1.000 \\
    80 & 1.000& 1.000 &1.000\\
    100 & 1.000  &1.000 &1.000\\

    \hline
    \end{tabular}}

    \label{tab:sc2}
\end{table}

\begin{table}[]
\caption{Empirical power for Scenario 3  at 5 \% significance level}
    \centering
    \begin{tabular} {cccccccccc}
    \hline
   $n$ & ($\lambda=2, \alpha=2)$ & ($\lambda=3, \alpha=2)$& ($\lambda=4, \alpha=2)$\\
\hline

   20 & 1.000 &1.000&1.000\\
    40 & 1.000  &1.000&1.000\\
    60 & 1.000  &1.000&1.000\\
    80 & 1.000 &1.000&1.000\\
    100 & 1.000  &1.000&1.000\\

    \hline
    \end{tabular}

    \label{tab:sc3}
\end{table}

\section{Data analysis}
As discussed before, the test is useful in the comparison of taxable incomes in income studies. Motivated by this, we conduct a data analysis to compare income from two states of India. The states of Kerala and Bihar were chosen for the study based on their distinct levels of income distribution resulting from prevailing socio-economic conditions.   The Consumer Pyramids Household Survey (CPHS) of the Centre for Monitoring Indian Economy (CMIE) is the source of the household-level income data for each state (the data available from  https://consumerpyramidsdx.cmie.com). It is a frequent, comprehensive survey that is conducted regularly to obtain data about Indian household demographics, spending, assets, and attitudes. Every year, three waves of data are collected, each lasting four months. We used data from Wave 28, which comprises information gathered between January and March 2023, for the analysis.

\begin{figure}[htbp]
\centering

\vspace{0.2in}
\begin{tabular}{c c }
     \includegraphics[width=6.2 cm]{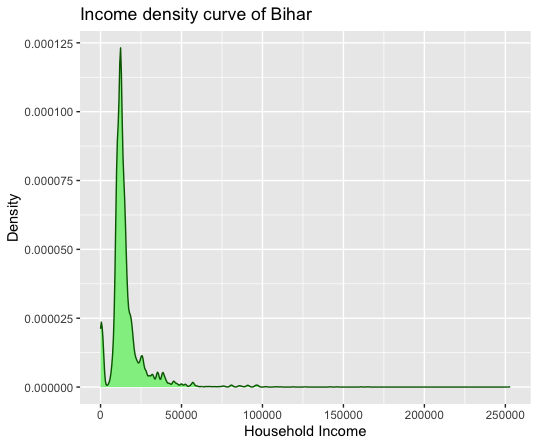}

 &\includegraphics[width=6.2 cm]{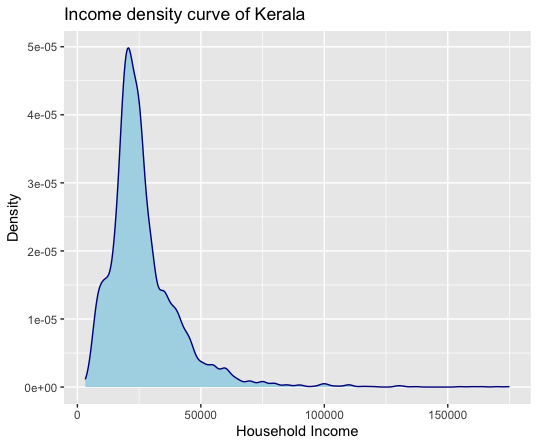}
\end{tabular}
\label{fig:den}
\caption{Income distribution of different states}
\end{figure}

\begin{table}[]
    \centering
     \caption{Descriptive statistics}
    \begin{tabular}{cccccccc}
    \hline
       & Kerala & Bihar \\
     \hline

      $n$   & 4310 & 7475\\
      \textit{Mean}  & 26829.06 &15713.65\\

      \textit{SD}  & 15185.06 & 12018.63 \\
      \textit{Range} & 171800 & 253000\\
      \textit{Skewness} & 2.75 & 4.69 \\
     \textit{Kurtosis}  &14.4 &43.5\\
      \hline
    \end{tabular}
    \label{tab:des}
\end{table}
The income distribution for each state is presented in Figure \ref{fig:den}.  From Figure  \ref{fig:den}, based on the distributional pattern, it can be deduced that there is a disparity in the income distribution among each state.  To further understand the differences in the distributional pattern between the states we have reported the descriptive statistics in Table  \ref{tab:des}.
Consequently, we anticipate unequal semivariance between these two states.

 To validate the claim, we performed the proposed test on the data. The calculated value of the JEL ratio  $ -2 \log R(\Delta) = 2296.14$. The result suggests that we can reject the null hypothesis at 5\% as well as 1\% significance level.  Thus, we can conclude that both states have unequal semivariance and have different income distribution.

\section{Concluding Remarks}
We proposed a test for testing the equality of the upper semivariance of two different independent populations. We employed the JEL ratio test for testing the same due to the implementation of the test as the usual normality-based test was difficult. The simulation study revealed that the proposed test has a well-controlled type I error and has good power for various alternatives. The real data analysis was also conducted to illustrate the use of the test and the results of the test were well in line with the conclusion drawn from the visual interpretation of the density plots which validates the utility of the proposed test. Right-censored observations are common in income studies (Jenkins et al., 2011, Kattumannil et al., 2021). One can develop a test for equality of upper semivariance under the right-censored case.

 While standard deviation and variance provide measures of risk, semivariance only looks at the negative fluctuations of an asset. By neutralizing all values above the mean, or an investor's target return,  semivariance estimates the average loss that a portfolio could incur. A similar test can be devised for the lower partial moment which is a  useful tool in such scenarios as it provides a measure for the downside risk.  Also, devising such a test can help in comparing the average downside risk between two different scenarios.

 Recently, Wu et al.  (2023) obtained a general expectation identity and discussed its uses in finding higher-order moments. Using this general identity, one can further study the properties of lower and upper-order partial moments.


\end{document}